\documentclass[12pt]{article}
\usepackage{epsfig}

\def\beq{\begin{equation}}
\def\eeq#1{\label{#1}\end{equation}}
\def\eeqn{\end{equation}}

\def\beqa{\begin{eqnarray}}
\def\eeqa#1{\label{#1}\end{eqnarray}}
\def\eeqan{\end{eqnarray}}

\let\bar=\overbar

\def\Dslash{\not{\hbox{\kern-4pt $D$}}}
\def\dslash{\not{\hbox{\kern-2pt $\del$}}}

\def\msb{{\bar{\ssstyle M \kern -1pt S}}}

\def\Title#1{\begin{center} {\Large {\bf #1} } \end{center}}

\begin{document}

\Title{Electroweak and QCD Results from the Tevatron}

\bigskip\bigskip

\begin{raggedright}  

{\it Dr. Junjie Zhu\index{Reggiano, D.}\\
Department of Physics\\
University of Michigan\\
Ann Arbor, MI, 48109, USA}
\bigskip\bigskip
\end{raggedright}

\section{Introduction}
The Tevatron collider has been remarkably successful and has so far delivered more than 11 fb$^{-1}$ 
of data to both the CDF and D0 experiments. Though the LHC has replaced the Tevatron as 
the world's most powerful collider, years of detector calibration, the huge size of the dataset 
and the nature of $p\bar{p}$ collisions will keep 
the Tevatron competitive in many selected topics in the near future.  
More than 10 fb$^{-1}$ of data has been collected by each experiment. Good understanding of the 
detector performance has been demonstrated by the high precision $W$ boson mass ($\Delta M_W=31$ MeV)
and top quark mass ($\Delta M_{t}=1.06$ GeV) measurements. 

We report the latest electroweak and QCD results from both experiments. 
Most analyses presented here used 4 - 6 fb$^{-1}$ of data.

\section{Electroweak Physics}
Large production cross sections of single $W$ and $Z$ bosons at the Tevatron allow precise 
measurements of the properties of intermediate weak bosons and determinations of fundamental 
parameters such as $M_W$ and $\sin^2 \theta_W$ etc. Studies with dibosons 
($V\gamma$, $VV'$ with $V, V'=W$ or $Z$) can provide better understanding of the diboson production 
processes and extraction of anomalous triple gauge couplings (aTGCs). 

The presence of both vector ($g_V^f$) and axial-vector ($g_A^f$) couplings of fermions to the $Z$ boson 
gives rise to an asymmetry in the distribution of the polar angle $\theta^*$ of the 
negatively charged lepton relative to the incoming quark direction in the rest frame 
of the decayed lepton pair. 
The forward-backward charge asymmetry, $A_{FB}$, 
is defined by $A_{FB} = (\sigma_F - \sigma_B)/(\sigma_F + \sigma_B)$, where $\sigma_F(\sigma_B)$ is 
the cross section for process with $\cos \theta*>0$ ($\cos \theta*<0$). 
The couplings $g_V^f$ and $g_A^f$ depends on the charge and isospin of the fermion, 
and $\sin^2 \theta_W$. 
At the Tevatron, the $Z/\gamma^*$ bosons are mainly produced via $u\bar{u}$ ($d\bar{d}$) annihilation. 
A precise measurement of the $A_{FB}$ distribution thus can be used to 
extract $\sin^2 \theta_W$ and the $Z-$light quark couplings ($g_V^u$, $g_A^u$, $g_V^d$, $g_A^d$).  
About 5 fb$^{-1}$ of data collected by the D0 experiment are used to measure the $A_{FB}$ distribution using 
$Z/\gamma^* \rightarrow ee$ events~\cite{D0_afb}. 
The $Z/\gamma^*$ candidates are selected using the requirement of 
two isolated electrons in the fiducial calorimeter region. 
The number of selected forward (backward) events is corrected for detector acceptance and selection efficiency 
in order to get the inclusive cross section for forward (backward) process. The $A_{FB}$ are measured in 15 
$M_{ee}$ bins in the range $50<M_{ee} < 1000$ GeV. 
Good agreement has been found between the measured $A_{FB}$ distribution and two theoretical predictions using {\sc pythia} and {\sc zgrad2}. 
The individual $Z-u(d)$ couplings are extracted by comparing the 
measured $A_{FB}$ distribution to templates generated with {\sc resbos} for different values of the $Z$-light quark 
couplings. 
World's best direct measurements of the $Z-u(d)$ couplings have been achieved. 
Events in the $Z$ pole region are also used to determine $\sin^2 \theta_W = 0.2309 \pm 0.0008 (stat.) \pm 0.0006 (syst.)$.

A precise measurement of the distribution of $\phi_{\eta}^*$ 
has been performed using 7.3 fb$^{-1}$ of data collected by the D0 detector~\cite{D0_zpT}. 
The variable, $\phi_{\eta}^*$, is defined as $\phi_{\eta}^* = \tan(\phi_{acop}/2) \sin(\theta^*_{\eta})$, 
where $\phi_{acop}$ is the acoplanarity angle, given by $\phi_{acop}=\pi - \Delta \phi^{\ell\ell}$, 
and $\Delta \phi^{\ell\ell}$ is the difference in azimuthal angle between the two lepton candidates. The 
variable $\theta^*_{\eta}$ is defined as $\cos(\theta_{\eta}^*)=\tanh[(\eta^- - \eta^+)/2]$, where 
$\eta^-$ and $\eta^+$ are the pseudorapidities of the negatively and positively charged leptons, respectively. 
The $\phi^*_{\eta}$ probes the same physical effects as the $Z/\gamma^*$ boson transverse momentum, 
but is less susceptible to the effects of experimental resolution and efficiency. 455k dielectron events 
and 511k dimuon events are used. The observed $\phi^*_{\eta}$ distribution is corrected for experimental 
effects using $Z/\gamma^*$ simulated MC events. The corrected particle level $\phi^*_{\eta}$ distributions 
are compared with the {\sc resbos} predictions in two (three) $Z/\gamma^*$ rapidity bins, 
using dimuon (dielectron) events. 
Predictions from {\sc resbos} are found to be unable to describe the detailed shape of the corrected data,
 and a prediction that includes the effect of small-$x$ broadening is strongly disfavored. 

A first measurement of the angular distributions of final state electrons in $Z/\gamma^* \rightarrow ee$ events 
has been performed by the CDF experiment using 2.1 fb$^{-1}$ of data~\cite{CDF_angular}. 
The angular distributions (the polar $\theta$ and azimuthal $\phi$ angles in the Collins-Soper frame) are 
studied and compared with predictions from several MC event generators. Good agreement has been observed for 
dependencies of four angular coefficients as a function 
of the dilepton transverse momentum. The relations of these angular coefficients are found to agree with the Lam-Tung relation,  
which is based on a spin-1 description of the gluon. This study demonstrates that at high 
values of the transverse momentum, $Z$ bosons are produced via $q\bar{q}$ annihilation and quark-gluon Compton processes. 

A $W\gamma$ cross section measurement has been performed in muon channel using 4.2 fb$^{-1}$ of data collected by the D0 
detector~\cite{D0_Wgamma}. The cross section times branching ratio for the process $W\gamma \rightarrow \mu\nu\gamma$ is 
determined for $p_T^\gamma>8$ GeV and $\Delta R_{\mu\gamma}>0.7$ 
The $\mu\nu\gamma$ events are selected by requiring one isolated muon and photon with large missing transverse 
energy. Photons are required to not be spatially matched to tracker activity. 
An artificial neural network discriminant is used to separate jets from photons. The dominant background is due 
to $W+$jet process, its contribution is estimated using a data-driven method. 
The $W\gamma$ cross section is measured to be $15.2 \pm 0.4 (stat) \pm 1.6 (syst)$ pb, which 
agrees with the SM expectation of $16.0 \pm 0.4$ pb. The photon $p_T$ distribution is used 
to set limits on the anomalous $WW\gamma$ couplings. The one dimensional 
95\% C.L. limits are $-0.14 < \Delta \kappa_{\gamma} < 0.15$ and $-0.02 < \lambda_{\gamma} < 0.02$. 

A study of the invariant mass distribution of two jets produced in association with a 
$W$ boson has been reported by the CDF experiment using 4.3 fb$^{-1}$ of data~\cite{CDF_Wjj}. 
Events are selected with one isolated electron or muon, large missing transverse energy 
and exactly two jets reconstructed with a fixed-cone algorithm with radius $\Delta R=0.4$.  
The dominant background due to $W+$jets and $Z+$jets processes are estimated using MC event generators and 
a {\sc geant}-based CDF detector simulation. The generators used are {\sc alpgen} with an interface to 
{\sc pythia} providing parton showering and hadronization. 
The observed distribution has an excess in the 120-160 GeV mass range which is not 
described by current theoretical predictions within the statistical and systematic uncertainties.  
Assuming only background contributions, the probability to observe an 
excess larger than in the data is $7.6 \times 10^{-4}$ corresponding to a significance of 3.2 standard deviations 
from a Gaussian distribution.

Both CDF and D0 experiments have measured the production cross section of $\sigma(p\bar{p} \rightarrow ZZ)$ with four 
charged lepton final state $\ell^+\ell^-\ell'^+\ell'^-$ ($\ell,\ell'=e$ or $\mu$)~\cite{D0_ZZ, CDF_ZZ1}. Ten candidate events are observed by 
the D0 experiment with an expected background of $0.37 \pm 0.13$ events. Fourteen candidates events are observed by the 
CDF experiment with an expected $ZZ$ signal of 10.4 events. The cross section is found to be 
$1.40 ^{+0.43}_{-0.37} (stat) \pm 0.14 (syst)$ pb (D0) and $2 \pm 0.58 (stat) \pm 0.32 (syst) \pm 0.12 (lumi)$ pb. 
CDF also performed a $ZZ$ cross section measurement using 
$ZZ \rightarrow \ell\ell\nu\nu$ events~\cite{CDF_ZZ2}. Larger branching ratio and backgrounds are expected compared with 
the four lepton channel. A neural network discriminant is trained to separated the $ZZ$ signal from the dominant Drell-Yan 
background. The cross section is measured to be $1.45^{+0.45}_{-0.42}(stat)^{+0.41}_{-0.30}(syst)$ pb.
The results from all three measurements are in agreement with a NLO prediction of $1.4 \pm 0.1$ pb. 

\section{QCD Physics}
Studies with inclusive jets, jets in association with a vector boson ($V+$jets), and photon pair ($\gamma\gamma$) provide powerful probes of the 
dynamics of hard QCD interactions. These studies can be used to make precision tests of QCD calculations in different regions 
of parton momentum fraction $x$ and hard-scattering scales $Q^2$, determine strong coupling constant $\alpha_s$, provide 
stringent constraints on PDFs, test and tune phenomenological models. 
The $V+$jets and $\gamma\gamma$ processes are also major backgrounds to many new phenomena searches. A better understanding 
of their production mechanisms will be useful to improve the new physics search sensitivities.

A first measurement of the inclusive three-jet differential cross section as a function of the invariant mass of the three jets 
($d\sigma_{3jet}/dM_{3jet}$) has been performed by the D0 experiment~\cite{D0_3jet}. 
The three-jet cross section is directly sensitive to the perturbative QCD (pQCD) matrix element of $O(\alpha_s^3)$, and therefore has a higher 
sensitivity to $\alpha_s$ as compared to inclusive jet and dijet cross sections, while having a similar sensitivity to the PDFs. 
Jets are defined by the Run II Midpoint jet algorithm with a cone of radius $\Delta R=0.7$. 
The measurement is made in different rapidity regions and for different jet transverse momentum requirements. 
The data are compared to theoretical predictions which have been obtained from NLO pQCD calculations with non-perturbative corrections 
applied. The non-perturbative corrections (including corrections due to hadronization and underlying event) are determined using {\sc pythia} 
with ``tune DW''. The best description of the data is obtained for the MSTW2008NLO and NNPDFv2.1 PDF parameterizations which describe both 
the normalization and the shape of the observed $M_{3jet}$ spectra. The PDF parameterizations from ABKM09NLO give a reasonable 
description of the data, although with a slightly different shape of the $M_{3jet}$ spectrum. The central results from the CT10 and HERAPDFv1.0 PDF sets 
predict a different $M_{3jet}$ shape and are in poorer agreement with the data.  

A first study of the substructure of jets with transverse momentum greater than 400 GeV has been performed by the CDF experiment using 6 fb$^{-1}$ of data~\cite{CDF_jetstructure}. 
Massive boosted jets constitute an important 
background in searches for various new physics models, the Higgs boson, and highly boosted top quark pair production. Jets are reconstructed 
using the Run II Midpoint cone algorithm using the {\sc fastjet} program and the anti-$K_T$ algorithm. 
The distributions of the jet mass, angularity, and planar flow are measured and compared with the pQCD calculations. 
Good agreement between {\sc pythia} predictions, 
the NLO QCD jet function predictions, and the data for the jet mass distribution above 100 GeV for Midpoint and anti-$k_T$ jets has been observed. The 
Midpoint and anti-$k_T$ algorithms are found to have very similar jet substructure distributions for high mass jets. Results also show that high mass 
jets coming from light quark and gluon production are consistent with two-body final states from a study of the angularity variable, and that further 
rejection against high mass QCD jets can be obtained using the planar flow variable.  

D0 experiment used 4.2 fb$^{-1}$ of data and measured the inclusive and differential cross sections of $W(\rightarrow e\nu)+n$ jet process ($n=1-4$)~\cite{D0_Wjet}. 
The background-subtracted data are corrected for detector resolution effects using a regularized inversion of the resolution matrix as implemented in the 
program {\sc guru}, with ensemble testing used to derive statistical uncertainties and unfolding biases. 
Considerably smaller uncertainties on $W+$jets production cross sections than previous measurements are achieved. 
The differential cross section is measured as a function of jet transverse momentum for jet multiplicities $n=1-4$, normalized to the inclusive $W \rightarrow e\nu$ 
cross section. The measurements are compared to NLO pQCD calculations ({\sc rocket+mcfm} and {\sc blackhat+sherpa}) in the $n=1-3$ jet multiplicity 
bins and to LO pQCD calculations in the 4-jet bin. The measurements are generally in agreement with 
pQCD predictions, although certain regions of phase space are identified where the calculations could be improved. 

CDF experiment used 8.2 fb$^{-1}$ of data and measured the inclusive and differential cross sections of 
$Z(\rightarrow \ell\ell)+ \ge n$ jet process ($\ell=e$ or $\mu$ and $n=1-3$)~\cite{CDF_zjets}. Jets are reconstructed using the 
Run II Midpoint algorithm with a cone of radius $\Delta R=0.7$. The measurements are corrected to the hadron level and compared to 
NLO pQCD predictions for $n=1, 2$ and LO pQCD predictions for $n=3$. The theoretical predictions have 
non-perturbative corrections applied. Good agreement for both the normalization and shape 
between data and theoretical predictions has been found for $n=1, 2$. For $n=3$, good agreement for the shape has 
been found between data and the LO predictions.  
D0 experiment performed a measurement of the ratio of the cross section 
of a $Z$ boson and at least one $b$-quark jet to the inclusive $Z+$jet cross section~\cite{D0_zbjets}. 
The $Z+b$ jet(s) process is a major background to the SM Higgs boson search in $ZH(H \rightarrow b\bar{b})$ associated 
production. This process is also sensitive to the $b$-quark PDFs. 
A discriminant that exploits the properties of the tracks associated to the jet is used to 
separate $Z+b$ jet(s) events from $Z+$ charm and light jet(s) events. 
The measurement ratio is $0.0193 \pm 0.0027$ for events having a jet with transverse momentum greater than 20 GeV and 
pseudorapidity less than 2.5, which is the most precise to date and is consistent with theoretical predictions of 
$0.0192 \pm 0.0022$. The dominant experiment uncertainties come from the uncertainty on the discriminant efficiency 
and on the shape of the discriminant templates used for the extraction of the $b$ jet fraction. 
Other important sources of uncertainty are the $b$ tagging efficiency,
the $b$ jet energy scale and reconstruction efficiency. 

D0 experiment measured the differential cross sections of direct photon pair process using 4.2 fb$^{-1}$ of data~\cite{D0_diphoton}. 
The direct photon pair production constitutes a large and irreducible background to searches for the SM Higgs boson 
search in diphoton channel and also new phenomena searches such as new heavy resonances and extra spatial dimensions. 
A requirement on the photon isolation distribution is used to suppress the 
contributions from photons produced in the fragmentation processes. The dominant backgrounds from $\gamma+$jet and jet$+$jet 
are estimated using a data-driven $4 \times 4$ matrix method. The single differential cross sections are measured as a function of the 
diphoton mass, the transverse momentum of the diphoton system, the azimuthal angle between the photons, and the polar 
scattering angle of the photon. In addition, double differential cross sections considering the last three kinematic variables 
in three diphoton mass bins are also measured. The results are compared with different pQCD predictions and event generators 
as {\sc resbos}, {\sc diphox} and {\sc pythia}, showing the necessity of including higher order corrections beyond NLO as well 
as the resummation to all orders of soft and collinear initial state gluons. Similar measurement has been performed by the CDF
experiment using 5.4 fb$^{-1}$ of data~\cite{CDF_diphoton}. The overall systematic uncertainty is limited to about 30\%. Similar conclusions 
have been made. Results also show that the inclusion of photon radiation in the initial and final states can significantly improve the 
{\sc pythia} parton shower calculation. 

\section{Conclusion}
Recent electroweak and QCD results from the Tevatron experiments are presented here. These measurements allow us to have better
understanding of the electroweak and QCD processes at hadron colliders. Most analyses shown here used 4-6 fb$^{-1}$ of data. 
With more than 10 fb$^{-1}$ collected by each experiment, we expect to see more individual and combined results from two Tevatron experiments.


\end{document}